\documentclass[useAMS,usenatbib]{mn2e}

\usepackage{graphicx,times,multirow,amsmath,bbold}

\def\src{RX J1856.5-3754}

\title[Polarized thermal emission from
X--ray Dim Isolated Neutron Stars:
the case of RX J1856.5-3754]{Polarized thermal emission from
X--ray Dim Isolated Neutron Stars:
the case of RX J1856.5-3754}
\author[D. Gonz\'alez Caniulef et al.]{D.
Gonz\'alez Caniulef$^{1}$\thanks{E-mail: 
denis.caniulef.14@ucl.ac.uk; s.zane@ucl.ac.uk; taverna@pd.infn.it;
turolla@pd.infn.it; kinwah.wu@ucl.ac.uk}, S.
Zane$^{1}$, R. Taverna$^{2}$, R. Turolla$^{1, 2}$, and K. Wu$^{1}$ \\
$^{1}$Mullard Space Science Laboratory, University College London, Holmbury St.
Mary, Dorking, Surrey, RH5 6NT, UK\\
$^{2}$Department of Physics and Astronomy, University of Padova, via
Marzolo 8, 35131 Padova, Italy}

\begin{document}

\date{}


\maketitle

\begin{abstract}
The observed  polarization properties of thermal radiation  from isolated,
cooling neutron stars depend on both the emission processes at the surface and
the effects of the magnetized vacuum which surrounds the star. Here we
investigate the polarized thermal emission from X--ray Dim Isolated Neutron
Stars, taking RX J1856.5-3754 as a representative case. The physical conditions
of the star outermost layers in these sources is still debated,  and so  we
consider emission from a magnetized atmosphere and a condensed surface, 
accounting for the effects of 
vacuum polarization as the radiation propagates in the star magnetosphere.
We have found that, for a significant range of viewing geometries, measurement of
the phase-averaged polarization fraction and phase-averaged polarization angle
at both optical and X--ray wavelengths  allow us to determine whether this
neutron star has an atmosphere or a condensed surface. Our results may
therefore be relevant in view of future developments of soft X--ray
polarimeters.
\end{abstract}

\begin{keywords}
Polarization ---
Radiation mechanisms: thermal --- stars: neutron --- X--rays: stars. 
\end{keywords}

\section{Introduction}
\label{intro}

X--ray dim isolated neutron stars (XDINSs), also known as the
``Magnificent Seven'', are a class of isolated, radio-silent X--ray
pulsars with peculiar properties, originally discovered by the
{\em ROSAT} satellite \cite[see e.g.][for a review]{turolla09}.
X--ray timing analysis allowed to measure the spin periods of all
sources \cite[$P \sim 3-12$~s; the latest addition being RX
J1605.3+3249, $P=3.39$ s;][]{pires14}, together with the period
derivatives, $\dot P\sim 10^{-14}-10^{-13} \rm{s~s}^{-1}$. These
translate into spin-down magnetic fields $B\sim
10^{13}$ -- $10^{14}$~G and characteristic ages $\tau_c$ of a few
Myr. When available, kinematic age estimates based on the
back-tracing of the star trajectory are typically shorter, $\sim
0.5$~Myr, and in agreement with those derived from the star
cooling history \cite[e.g.][and references therein]{mignani13}.

XDINSs are quite close sources, possibly cradled in the young
stellar clusters forming the Gould Belt \cite[][]{popov03}. The
distances estimated from the hydrogen column density are $\la
500$~pc \cite[][]{posselt08} and parallax measurements for \src\
and RX J0720.4-3125 provide values of 123 and 360 pc, respectively
\cite[][and references therein]{walter10,kvk09}.

The Seven exhibit a purely thermal spectrum at X--ray energies with
no evidence for a high-energy, power-law component often detected
in other isolated NS classes. The X--ray luminosity is
$10^{31}$ -- $10^{32}\ {\rm erg\, s}^{-1}$, fully consistent with
surface blackbody emission with temperatures $\sim 40$--100 eV and
(radiation) radii of a few kilometers, as derived from X--ray
spectral fits \cite[see e.g.][]{kvk09,turolla09}. With the only
exception of \src,  broad absorption features have been found in
all XDINSs. These features have energies $\sim 300$ -- 700 eV,
equivalent widths of $\sim 50$ -- 150 eV and, as in the case of RX
J0720.4-3125, may be variable. In the latter source a second,
strongly phase-dependent line was very recently reported
\cite[][]{borghese15}.

Optical counterparts, with magnitudes $\ga 25$, have been
identified (to a varying degree of confidence) for all the Seven
on the basis of proper motion measurements or positional
coincidence \cite[e.g.][]{turolla09, kaplan11}. The optical flux
appears to exceed the extrapolation of the X--ray blackbody at low
energies by a factor $\sim 5$ -- $50$ and deviations from a
Rayleigh-Jeans distribution have been reported in some sources
\cite[notably RX J2143.0+0654;][]{kaplan11}.

The nature of the surface emission from XDINSs is still a debated
issue. According to the conventional picture, isolated, cooling
neutron stars are covered by an atmosphere which reprocesses the
thermal radiation coming from the outermost stellar layers
\cite[see e.g.][for a review]{potekhin14}. More recently, it has
been appreciated that the low surface temperature ($\la 100\ {\rm
eV}$) and the strong magnetic field ($\ga 10^{13} \ \rm G$) of
XDINSs may produce a phase transition in the surface layers,
leaving a bare neutron star with a condensed (either solid or
liquid) surface \cite[][see also \citealt{turolla09},
\citealt{potekhin14}]{laisalp97,lai01,burwitz03,turolla04,medinlai07}.

The merits of the two models in explaining the observed
multiwavelength spectral energy distribution (SED) of the XDINSs
have been assessed in several studies, which are, however,
hampered by the present poor knowledge of the star internal
magnetic field structure and hence of the surface temperature
distribution. Mid-Z element atmosphere models have been proposed
for RX J1605.3+3249 \cite[][]{mori07}. \cite{pazor06} and
\cite{ho07} used the condensed surface emission model to explain
the observed properties of RX J0720.4-3125 and \src, respectively,
although in the latter a thin, magnetized, H atmosphere on top of
the condensate was added \cite[as originally suggested by][see
also \citealt{zane04}]{motch03}. A detailed, comparative
investigation of atmospheric/condensed surface emission models was
presented by \cite{suleim10}, who also accounted for the possible
presence of a thin H atmosphere around the condensed surface.
Results were then applied to fit the SED of RX J1308.6+2127,
showing that for this source an interpretation in terms of
emission from a condensed surface with a thin atmosphere is
favored \cite[][]{hambar11}.

Polarimetric measurements both at optical and X--ray energies can
provide a valuable tool to better understand the physical
properties of the neutron star surface. Current 8-m class
telescopes, e.g. the VLT, already allow to perform polarization
measures for faint sources like the XDINSs. X--ray polarimetry
missions are at an advanced stage of development. The X--ray
Imaging Polarimetry Explorer\footnote{http://www.isdc.unige.ch/xipe/}(XIPE),
the Imaging X--ray Polarimetry
Explorer\footnote{\citet[][]{2013SPIE.8859E..08W}}(IXPE), and the Polarimeter
for
Relativistic Astrophysical X--ray
Sources\footnote{\citet[][]{2015AAS...22533840J}}(PRAXyS) have been selected for
the
study phase of the ESA M4 and the NASA SMEX programmes. They will
open the possibility to perform X--ray polarimetry and pave the way
toward the construction of a X--ray polarimeter efficient in the
soft X--rays.

Radiation from the surface of a neutron star is expected to be
intrinsically polarized, because the strong magnetic field
introduces an anisotropy in the medium in which electromagnetic
waves are propagating. This, in turn, causes the opacity of the
two normal modes (the ordinary and extraordinary) to be different,
so that the emergent radiation carries a net polarization \cite[see
e.g.][]{hardlai06}. The
expected polarization pattern is different whether emission comes
from an atmosphere \cite[][]{vanadel06} or a condensed surface
\cite[][]{potekhin12}.
Thus, the study of the
polarized emission from XDINSs can give us insight about the
nature of the surface of strongly magnetized NS and ultimately probe the
properties of the matter under strong magnetic fields.

Among the ``Magnificent Seven", the most promising source for the
study of polarized emission in the optical and X-ray band is \src \
(hereafter RX J1856). This is the brightest and nearest XDINS,
with $V=25.58$, a nearly $\lambda^{-4}$ optical-UV SED
\cite[][]{vkk01, kaplan11} and a X--ray spectrum well modeled by
two blackbody components \cite[$T^{\infty}_\mathrm c\sim 40$~eV
and $T^{\infty}_\mathrm h\sim 60$~eV\footnote{Here $T^{\infty}$
denotes the temperature measured by an observer at
infinity.};][]{sartore12}. {The period derivative
of RX J1856 has been obtained by \citet{vkk08}, $\dot P\sim 3\times 10^{-14}\
{\rm s/s}$, which translates into a  
spin-down magnetic field of $B\sim 1.5\times10^{13}$~G. 
An alternative estimate, 
$B_p = 6\times10^{12}$~G (at the magnetic pole, assuming a dipole 
model), has been derived from fitting continuum 
 models to the observed optical and X-ray spectrum \citep{ho07}. 
These relatively
strong magnetic fields imply that  a non vanishing
degree of polarization is indeed expected in the thermal emission of the
source.}

In this paper we derive expectations for the polarization observables, focusing
on
the case of RX J1856. First, we briefly summarize the theoretical brackground
to calculate the thermal emission from a magnetized atmosphere and a condensed 
surface (\S~\ref{theory}). Then,  we  proceed to the calculation of the
intrinsic polarization properties (i.e. those at the star surface)
for a magnetized, fully-ionized H atmosphere and for a condensed
surface (\S~\ref{results}). We then turn to the evaluation of the
polarization fraction and the polarizations angle as measured by a
distant observer, accounting for the effects of QED (vacuum
polarization) and of the non-uniform star magnetic field
(\S~\ref{vac}); this is done following closely the approach
described in \citet[hereafter paper I]{taverna15}. Results are
presented in \S~\ref{obspol}. Discussion and conclusions follow in
\S~\ref{disc}.

\section{Theoretical framework}
\label{theory}

In this section we briefly outline the basic physical inputs of
our model. Since XDINSs are slow rotators ($P\sim$ a few seconds, 
\citealt{turolla09}),
we neglect the effects of rotation and assume that the space-time
outside the star is described by the vacuum Schwarzschild
solution. Moreover, despite our treatment can handle general
axially-symmetric magnetic fields, in the following we restrict to
the case in which the neutron star field is dipolar, $\mathbf B =
B_\mathrm
p=(R_\mathrm{NS}/r)^3(f_\mathrm{dip}\cos\theta,g_\mathrm{dip}\sin\theta/2,0)$,
where $B_\mathrm p$ is the polar field strength,
$R_\mathrm{NS}$ is the star radius ($M_\mathrm{NS}$ denotes the
star mass) and $r$, $\theta$ are the radial coordinate and the
magnetic co-latitude, respectively. The two functions
\begin{flalign}
\arraycolsep=1.4pt\def\arraystretch{1.8}
\begin{array}{l}
f_\mathrm{dip} =-\dfrac{3}{x^3}\left[\ln(1-x)+\dfrac{1}{2}x(x+2)
\right] \\
g_\mathrm{dip} =\sqrt{1-x}\left(-2f_\mathrm{dip}+\dfrac{3}{1-x}\right)\,,
\end{array} & &
\end{flalign}
account for relativistic {corrections \cite[][]{ginzozer65,mustsy86}}, with
$x=R_\mathrm s/r$, and $R_\mathrm s=2GM_\mathrm{NS}/c^2$.

\subsection{Ray tracing method}
\label{ray}

Given an emission model characterized by a specific intensity $I_\nu$, which in
general depends
on the photon frequency $\nu$ and direction $\mathbf k$, and on the position on
the star surface,
the spectral and polarization properties at infinity are computed by summing the
contributions of the surface elements which
are into view at a given rotational phase.

Following \cite{zane06} and paper I,
we introduce the two
angles $\chi$ and $\xi$: the former is the angle between the
 line-of-sight (LOS, unit vector $\boldsymbol{\ell}$) and the spin axis
($\mathbf p$), while the latter is that  between the magnetic
(dipole) axis ($\mathbf{b}_{\rm dip}$), and the spin axis. We
further introduce a (fixed) coordinate system, $(X,Y,Z)$ with the
$Z$-axis parallel to $\boldsymbol{\ell}$ (i.e. along the LOS) and the
$X$-axis in the  $(\boldsymbol{\ell},\mathbf{p})$ plane, and a
co-rotating coordinate system, $(x,y,z)$, with the $z$-axis
parallel to $\mathbf{b}_{\rm dip}$ and the $x$-axis defined below.
The associated polar angles are  $(\Theta_\mathrm S, \Phi_\mathrm
S)$ and $(\theta,\phi)$, respectively. In the fixed frame, the
cartesian  components of $\mathbf p$ and $\mathbf{b}_{\rm dip}$
are  $\mathbf{p} = (\sin \chi, 0, \cos \chi)$ and $\mathbf{b}_{\rm
dip} = (\sin\chi\cos\xi - \cos\chi\sin\xi\cos\gamma,
\sin\xi\sin\gamma, \cos\chi\cos\xi + \sin\chi\sin\xi\cos\gamma)$,
where $\gamma = \omega t$ is the phase angle ($\omega=2\pi/P$,
and $P$ is the star rotational period). We also define an
additional vector,
$\mathbf{q}=(-\cos\chi\cos\gamma,\sin\gamma,\sin\chi\cos\gamma)$,
which is a unit vector orthogonal to $\mathbf{p}$ and rotating
with angular velocity $\omega$ (in the fixed frame). The $x$-axis
of the rotating coordinate system is then chosen in the direction
of the  component of $\mathbf{q}$ perpendicular to
$\mathbf{b}_{\rm dip}$,
\begin{equation}
\mathbf{q}_\perp =
\frac{\mathbf{q}-(\mathbf{b}_{\rm dip}\cdot\mathbf{q})\mathbf{b}_{\rm
dip}}{\left[
1-(\mathbf{b}_{\rm dip}\cdot\mathbf{q})^2)\right]^{1/2}}\,.
\end{equation}
The transformations linking the pairs of polar angles in the two systems are 
\cite[see][paper I]{zane06}
\begin{eqnarray}
\label{polartransform}
\cos\theta &=& \mathbf{b}_{\rm dip}\cdot\mathbf{n}\nonumber\\
\cos\phi &=& \mathbf{n}_{\perp}\cdot\mathbf{q}_{\perp}
\end{eqnarray}
where $\mathbf{n}= (\sin\Theta_\mathrm S\cos\Phi_\mathrm
S,\sin\Theta_\mathrm S\sin\Phi_\mathrm S,\cos\Theta_\mathrm S)$ is
the radial unit vector in the fixed coordinate system and
$\mathbf{n}_{\perp}$ is defined in analogy with
$\mathbf{q}_{\perp}$. Equations (\ref{polartransform}) are needed
to express the intensity, which is naturally written in terms of
the magnetic coordinate angles $(\theta\,,\phi)$, see \S
\ref{atmo}, \ref{conden}, in terms of the polar angles of the
fixed frame $(\Theta_\mathrm S\,, \Phi_\mathrm S)$ over which
integration is performed.

At each phase the monochromatic flux detected by an observer at distance $D\gg
R_\mathrm{NS}$ is
obtained by integrating the intensity (in the fixed coordinate system) over
the visible part of the surface \cite[see e.g.][paper
I]{page95,zane06}
\begin{equation}
\label{flux}
F_\nu(\gamma) =\left(1-\frac{R_\mathrm s}{R_\mathrm {NS}}\right)\frac{R_{\mathrm
{NS}}^2}{D^2}
\int_0^{2\pi}d\Phi_\mathrm S\int_0^1  
I_\nu(\mathbf k,\theta,\phi) 
du^2\,,
\end{equation}
where $u=\sin\bar\Theta$. The two angles, $\Theta_\mathrm S$ and $\bar\Theta$,
are
related by the ``ray tracing'' integral
\begin{equation}
\bar\Theta =
\int_0^{1/2}dv\, \sin\Theta_
\mathrm S
\left[\frac{1}{4}(1-x)-(1-2vx)v^2 \sin\Theta_\mathrm S^2 \right]^{-1/2}\, .
\end{equation}
For $x\ll 1$ Newtonian geometry is
recovered and $\bar\Theta = \Theta_\mathrm S$. 
A $50\times50$ mesh in $\cos\Theta_\mathrm S$ and $\Phi_\mathrm S$, equally spaced in the
$[0,1]$ and $[0,2\pi]$
intervals, respectively,  is typically used in our numerical calculations.

In the case of radiation (linearly) polarized in the two normal
modes (the ordinary, O, and the extraordinary, X, mode), the total
intensity is just the sum of the intensities in the two modes
\begin{equation}
I_\nu(\mathbf k,\theta,\phi) = I_{\nu,\mathrm O}(\mathbf k,\theta,\phi) +
I_{\nu,\mathrm X}(\mathbf k,\theta,\phi)
\end{equation}
and we define the ``intrinsic'' degree of polarization\footnote{Notice that
for a given surface element, the normal modes computed in our atmospheric or
crustal model are  defined with respect to a reference frame that depends on the
local  direction of the magnetic field.  To take into
account a varying magnetic field over  the star surface, a proper calculation of
the degree of polarization  requires a rotation
of the local reference frames (for the normal modes)  to the common reference
frame of a polarimeter (this will be
performed in  \S~\ref{vac}, see also \citealt{pavzav00}).}, i.e. that
at the source,
as
\begin{equation}
\label{polint}
\Pi_\mathrm L^\mathrm{EM}=\frac{F_\mathrm X-F_\mathrm O}{F_\mathrm X+F_\mathrm
O}\,,
\end{equation}
where $F_\mathrm{X,O}$ is the monochromatic, phase-dependent flux
in each mode, defined as in equation (\ref{flux}).

\begin{figure}
\center
\includegraphics[width=10.0cm, viewport=0 0 550
391,clip]{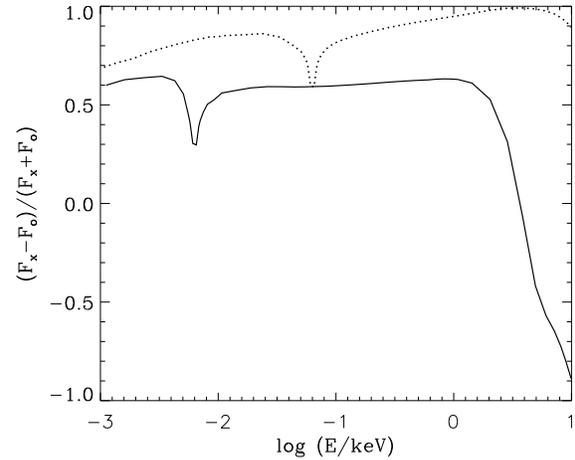}
\caption{Degree of polarization 
for emission from a pure H ionized atmosphere with
$T=10^{6.5}$~K and magnetic field perpendicular to the surface. The solid 
line corresponds to $B=10^{12}$~G, and the dotted line to $B=10^{13}$~G. 
See text for details.}
\label{fig:0pol}
\end{figure}

\begin{figure*}
\includegraphics[width=8cm]{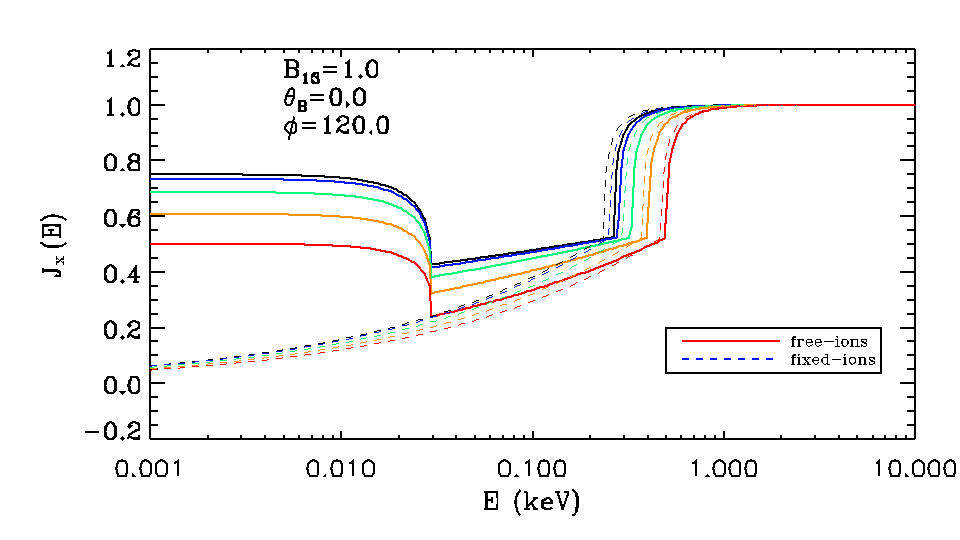}
\includegraphics[width=8cm]{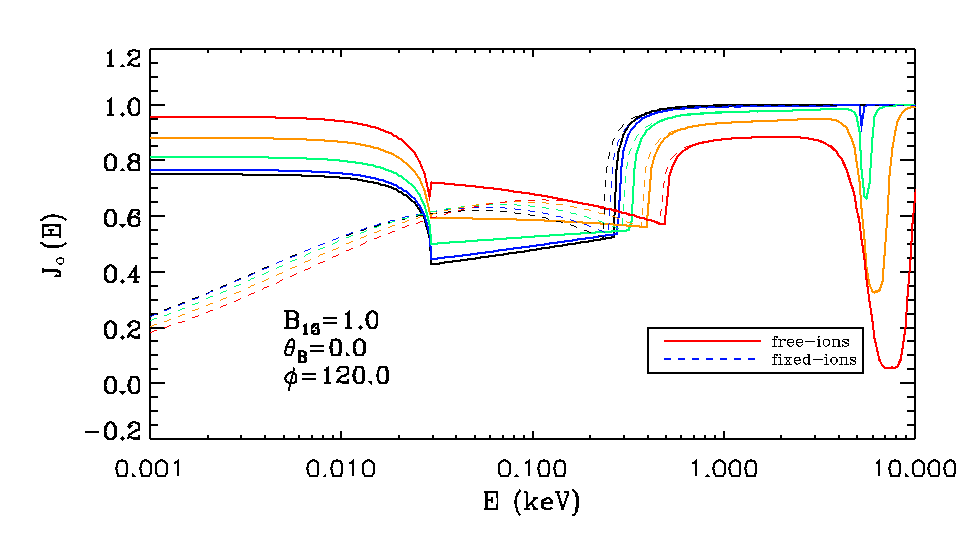}
\caption{The condensed-surface emissivity in the X (left) and O mode (right).
Full (dashed) lines refer to the free (fixed)-ions approximation. 
The different colours are for different values of $\theta_\mathrm k$ ($0^\circ$,
black;  $15^\circ$, blue; $30^\circ$, green;  $45^\circ$, orange; $60^\circ$,
red). Here
$\phi_\mathrm k=120^\circ$, $\theta_\mathrm B=0$ and $B=10^{13}$ G.}
\label{fig:condenemiss}
\end{figure*}

\subsection{Atmosphere}
\label{atmo}

Atmospheres around cooling NSs are commonly modeled by  considering
a gas in radiative and hydrostatic equilibrium. Since the scaleheight  of the
atmosphere, $h\sim kT/m_p g \sim 10$~cm, is much smaller than the star radius,
the radiative
transfer equation is solved in the  plane-parallel approximation, usually
assuming an atmosphere in local thermodynamic equilibrium. Atmospheric models
have 
been presented by a number of authors, under different assumptions and
accounting for different degrees of 
sophistication in the description of radiative processes and the plasma
composition 
\cite[see e.g.][for a review]{potekhin14}. 
Our objective in this work is to derive simple expectations for the difference in the 
polarization  signal emitted in the case the NS is covered by a gaseous layer or it is 
``bare'' (see \S~\ref{conden}). We therefore adopt the assumption that of a fully ionized pure H atmosphere 
and avoid the complication of atmospheric compositions. The emergent
intensity is computed using the 
numerical method presented in \citet[see also
\citealt{lloyd03b,zane06}]{lloyd03a}. 
The code exploits a complete linearization technique for solving the stationary 
radiative transfer equations for the two normal polarization modes in a
plane-parallel slab, by including 
the effect of the magnetic field inclination with respect to the surface normal.
The source term accounts for 
magnetic bremsstrahlung and magnetic Thompson scattering.  

The spectral calculations have four input parameters:  the (local) effective
temperature $T$ and magnetic field strength, $B$, the angle between the
local magnetic field $\mathbf B$ and the surface normal $\mathbf n$, 
$\theta_\mathrm B$, 
and the surface  gravity, $g$. The
code solves for the emergent intensity $I_\nu(\mathbf k) \equiv I(E, \mu_\mathrm
k, \phi_\mathrm k)$ 
for a range of photon
energies $E=h \nu$, photon co-latitudes and azimuthal angles relative to the 
slab
normal, $\theta_\mathrm k= \arccos (\mu_\mathrm k)$ and $\phi_\mathrm k$,
respectively. 
The $\phi_\mathrm k= 0$ direction
is defined by the projection of the magnetic field on the symmetry planes. We
should notice that, 
since the magnetic field
introduces an anisotropy in the opacities,  the  emergent intensity is not
symmetric with respect to a rotation around the surface normal 
but instead it depends on both, $\mu_\mathrm k$ and $\phi_\mathrm k$. 
For the particular case in which 
$\theta_\mathrm B  =0$, symmetry with respect to  $\phi_\mathrm k$ 
is restored, so the calculation is restricted to the $\mu_\mathrm k$-dependent
intensities. Moreover, even in the general case $\theta_\mathrm B 	>0$,
thanks to  the symmetry properties of the opacities,
the calculation of the emergent spectrum can be restricted to $0<\phi_\mathrm
k<\pi$.

Since we are considering photon energies well below the electron cyclotron
frequency, the opacity for 
O-mode photons is almost unaffected by the magnetic field, while that for X-mode
ones is substantially 
reduced \citep{hardlai06}. Therefore, in general, the emergent intensity of the
X-mode is much
larger than that of  the O-mode, 
resulting in an emergent  radiation with a non-null degree of polarization.
This is illustrated in
Figure \ref{fig:0pol},
that shows the intrinsic polarization fraction, as a function of the energy, 
for a single model, assuming a parallel magnetic field ($\theta_\mathrm B=0$).
As it can be seen, for 
$B= 10^{13}$~G, the polarization fraction is relatively high in the optical band ($\sim 70\%$)
and it tends to increase at high energies.

\begin{figure}
\center
\includegraphics[width=10.0cm, viewport=0 0 550
391,clip]{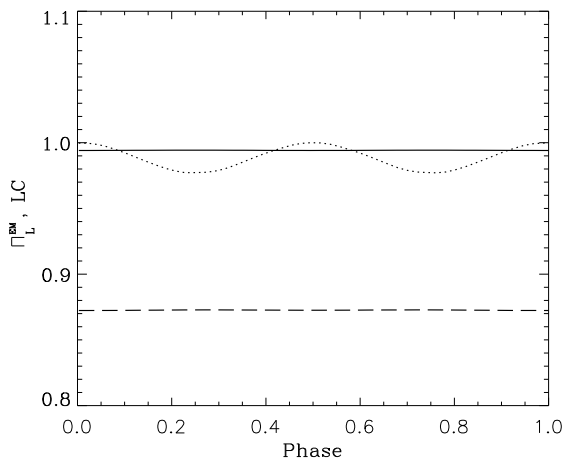}
\caption{Pure H, fully ionized atmosphere model for the emission from RX J1856. 
The dotted line correspond to the the X--ray 
lightcurve (LC), while the solid and dashed lines correspond to the phase 
dependent polarization fraction in the X--ray band and in the optical band, 
respectively. The viewing geometry is {such that $\chi=90^\circ$ and
$\xi=15^\circ$}. See text for details.}
\label{fig:0lc}
\end{figure}

\subsection{Condensed surface}
\label{conden}

Magnetic fields higher than $\sim m_e^2e^3c/\hbar^3 = 2.4\times 10^{9}\
\mathrm G$ 
change the properties of atoms,
confining electrons in the direction perpendicular to the magnetic
field. These elongated, cylindrical atoms can form molecular
chains by covalent bonding along the magnetic field lines. In
turn, chain-chain interaction can lead to the formation of a
condensed phase, as originally suggested {by \citet[][see also
\citealt{ruder71}]{laisalp97}}. The
cohesive energy for linear chains increases with the magnetic
field strength and it is expected that, for sufficiently strong
magnetic fields, there is a  critical temperature,
$T_\mathrm{crit}$, below which a phase transition  between a
gaseous and a condensed state occurs. This critical temperature
depends on composition and increases with magnetic field strength
\cite[][]{lai01,medinlai07}. Most recent estimates give
$T_\mathrm{crit} \approx 10^5 [5+2(B/10^{13}\ \rm G)]$~K  for Fe
composition, meaning that a phase transition may occur for surface
temperatures $T\ga 10^6$~K if the field is stronger than $\sim
10^{13}\ \rm G$. Since these are the typical surface temperature
and magnetic field found in XDINSs,  the thermal spectrum may
indeed come from a condensed surface \cite[][]{turolla04}.

\begin{figure*}
\includegraphics[width=7.2cm]{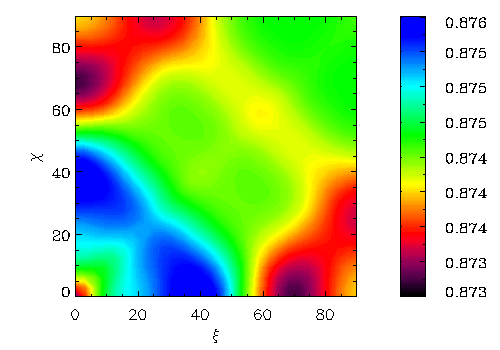}
\includegraphics[width=7.2cm]{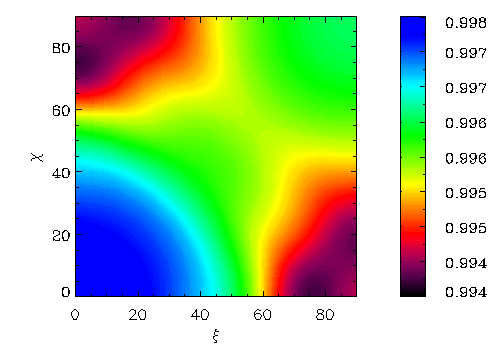}
\caption{Contour plots of the intrinsic, phase-average polarization fraction
in the ($\chi$, $\xi$) plane  for the gaseous atmosphere model, with the left
panel for the optical (B-filter) band and 
the right panel for  the X--ray (0.12 - 0.39 keV, at
infinity) band. See text for details.}
\label{fig:0cntAt}
\end{figure*}

The spectral properties of emission from a neutron star with a
condensed surface were investigated in several papers since the
pioneering work of \cite{brink80}. In essence, the intensity is
computed from the emissivity coefficient, $j_\nu$, which is in
turn related to the  reflectivity  via Kirchhoff's law. The latter
is calculated applying Snell's law at the interface between the
vacuum and the condensed phase. The boundary conditions for the
transmission of an electromagnetic wave across the two media give
the amplitude of  the reflected waves \cite[][see also
\citealt{potekhin14}]{turolla04,pazor05,vanadel05,suleim10}.

There are uncertainties in this kind of calculation. Our present knowledge of the 
condensate is poor, and the lacking of a reliable expression of the dielectric tensor 
hinders the correct derivation of the reflectivity. Previous works adopted a
simplified treatment, in which only the limits of ``free ions''
(no account for the effects of the lattice on the interaction of
the electromagnetic waves with ions) and of ``fixed ions'' (no ion
response to the electromagnetic wave because lattice interactions
are dominant) were considered. The true reflectivity of the
surface is expected to lie in between these two limits.

Here we maintain the same approach and use the analytical
approximations by \cite{potekhin12} to compute the emissivities in
the two normal modes. {They depend on the magnetic
field $B$, the photon direction $\mathbf k$ and energy, and the
angle between $\mathbf k$ and $\mathbf B$, $\theta_\mathrm{Bk}$. 
However, the modes $1,\, 2$ of \cite{potekhin12} are not 
defined with respect to the local direction of $\mathbf{B}$ but with respect to the local normal
$\mathbf{n}$, with mode 1 perpendicular to both $\mathbf{k}$ and $\mathbf{n}$.
As a consequence, the emissivities $j_{\nu,\mathrm i}$ ($\mathrm i =1,\,2$) do
not
coincide with those in the X and O modes, unless $\mathbf{n}$ and $\mathbf{B}$ are parallel, i.e. $\theta_\mathrm{B}=0$. The transformation linking the emissivities in the
two basis is given in Appendix B of \cite{potekhin12}\footnote{Note that
there is a typo in equations B.6, where the array in the left hand side should
be a matrix, and 
B.12, where $\cos^2\theta_\mathrm k - \sin^2\theta_\mathrm k$ should be $\cos^2\theta_\mathrm k + \sin^2\theta_\mathrm k=1$.}. Once the transformation is perfomed}
the intensity  of  the emergent  radiation in the X and O  modes
follows, by assuming the radiance of a blackbody, $B_\nu
(T)=2h\nu^3/[c^2 (\exp{(h\nu /kT)}-1)]$,

\begin{eqnarray}
&I_{\nu,\mathrm O}=j_{\nu,\mathrm O}(\nu, B,\mathbf k,\theta_\mathrm{Bk})B_\nu
(T)\nonumber\\
&I_{\nu, \mathrm X}=j_{\nu, \mathrm X}(\nu, B,\mathbf
k,\theta_\mathrm{Bk})B_\nu (T)\,.
\end{eqnarray}
Figure \ref{fig:condenemiss} shows the emissivity in the two normal modes,
calculated in the two limits (``free'' 
and ``fixed'' ions), for different values of $\theta_\mathrm k$.

\section{The model for RX J1856}
\label{model}
\begin{figure}
\center
\includegraphics[width=10.cm, viewport=0 0 550
391,clip]{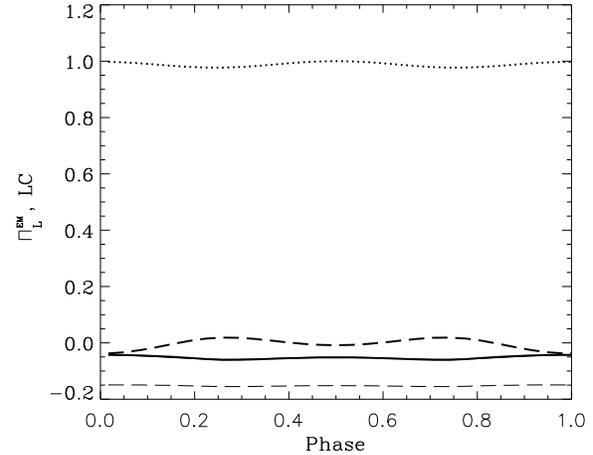}
\caption{Same as in Figure \ref{fig:0lc} for the condensed surface. 
The dotted line is the X--ray 
lightcurve, while the solid and dashed lines are the phase 
dependent polarization degree in the X--ray and in the optical band, 
respectively. Thick curves correspond to free ions, and thin curves to 
fixed ions. {Note that the X--ray lightcurve and phase dependent
polarization degree in the optical band are almost indistinguishable in the two
cases.}  Here $\chi=90^\circ$ and $\xi \approx 18^\circ$. See text for 
details.}
\label{fig:0lcCon}
\end{figure}
{In the following  we consider a NS with mass 
$M_\mathrm{NS}=1.5~M_\odot$ and
radius $R_\mathrm{NS}=12$~km, which is compatible with expectations 
from modern equations of state such as APR or BSk21 models
\citep{akmal98,goriely10}. The value of the radius is also in agreement with the
estimates derived by   \citet{sartore12} and \citet{ho07}, assuming a source
distance of  120 pc \citep{walter10}. This choice translates into a 
gravitational red-shift factor at the star surface $1+z = 1.26$}. The
rotational period of RX J1856 is $P=7$~s and the X--ray pulsed fraction is the
lowest among the XDINSs, $\sim 1.3\%$ \cite[][]{tiengo07}. {The polar strength
of the dipole field is taken to be 
 $B_\mathrm p = 10^{13}$~G, a 
value which is somehow intermediate between the spin-down measure 
and the estimates from spectral fitting \citep{vkk08, ho07}}. We
assume that the magnetic field is dipolar (see \S \ref{theory}) and that the
surface temperature distribution is that induced by the core-centered dipole.
Since for fields higher than $\sim 10^{11}\ \mathrm G$, electron conduction
across the field lines is strongly suppressed, the meridional temperature
variation is
$T_\mathrm{dip}\simeq T_\mathrm p\vert\cos\theta_\mathrm
B\vert^{1/2}$, where $T_\mathrm p$ is the polar value of the
temperature \cite[e.g.][]{green83}. {We checked that this simple 
expression 
for $T_{\mathrm dip}$ differs only slightly ($\la 6\%$) from the more 
accurate formula by \citet{potekhin15} for $\theta\la 80^\circ$.} 
However, taken face value, the previous expression for $T_\mathrm{dip}$
yields vanishingly small values near the magnetic equator. The
analysis of \cite{sartore12} has shown that the X--ray spectrum of
RX J1856 is best modeled in terms of two blackbody components
with $kT^\infty_\mathrm{h}\sim 60$~eV and
$kT^\infty_\mathrm{c}\sim 40$~eV. To account for this in a simple way, we
actually adopt a temperature profile given by $T_\mathrm 
s=\max(T_\mathrm{dip},T_\mathrm{c})$ with $ T_\mathrm
p=T_\mathrm{h}$, where $T_\mathrm{h,c}=T^\infty_\mathrm{h,c}/(1+z)$.

\subsection{Intrinsic polarization degree}
\label{results}
\subsubsection{Atmosphere}
\label{resatmo}
We first consider the case in which the star is surrounded by a gaseous
atmosphere. The star surface is divided in six angular patches in magnetic
colatitude,
centered  at the values $\theta=\{0^\circ, 10^\circ, 30^\circ, 50^\circ, 
70^\circ, 89^\circ \}$.
By using the magnetic and temperature profiles previously described we compute,
for each $\theta$, the 
local magnetic field strength, $B$, the angle $\theta_\mathrm B$ between the
magnetic field and the normal to the surface, 
and hence the temperature, $T$. 
We then compute a set of atmospheric models corresponding to the six $\theta$ 
angles. Since the models are computed using different
integration grids in the photon phase space \cite[because the choice of
the photon trajectories along which the radiative transfer is solved needs
to be optimized to ensure fast convergence
at the different values of magnetic field strength and inclination,
see][]{lloyd03a}, 
we  reinterpolate all model outputs over a common grid. 
This results in three 4-D arrays for the emergent intensity
$I^i_\nu(\mathbf k,\theta) \equiv  I^i(E,\mu_k,\phi_k,\theta)$ ($i=1,...3$)  
which contain the total intensity and the intensities for the ordinary and
extraordinary modes, respectively. 
{In order to take into account the emission from the southern magnetic
hemisphere of the star, we use the previous 4-D arrays with the 
substitutions $\theta_B \rightarrow \pi - \theta_B$ and $\phi_k 
\rightarrow \pi - \phi_k$, 
 which is justified by the simmetry properties of the opacities.}
By using the ray tracing method described in \S~2, we can then compute
lightcurves, phase resolved spectra and polarization fractions for each
choice of the angles $\chi$ and $\xi$.  
As an example, 
Figure \ref{fig:0lc} 
shows the X--ray lightcurve ($0.12-0.39$ keV) and the phase--dependent 
polarization degree in the X--ray and optical\footnote{The B--filter
is in the energy range  $2.5-3.1$~eV {at infinity}.} (B--filter) 
bands, for 
$\chi=90^\circ$ 
and $\xi=15^\circ$. For this particular viewing 
geometry,  
the X--ray pulsed fraction is $1\%$, in agreement 
with the observed data and, as illustrated in the figure, the 
polarization degree is expected to be substantial and constant in phase.
\begin{figure*}
\includegraphics[width=7.2cm]{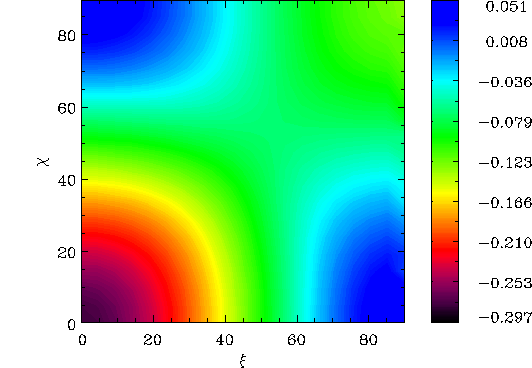}
\includegraphics[width=7.2cm]{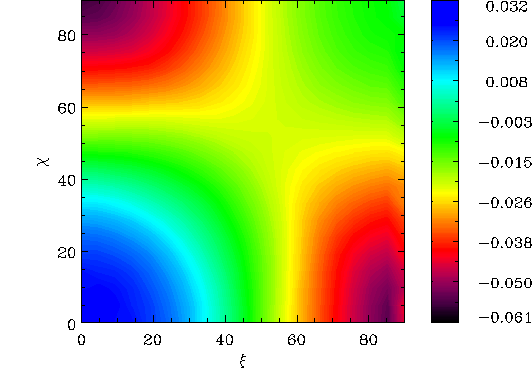}
\caption{Contour plots of the intrinsic, phase-average polarization fraction
in the  ($\chi$, $\xi$) plane  for the free-ion crustal emission model, with the
left panel for the optical (B-filter) band  and the right panel for  the X--ray
(0.12 - 0.39 keV) band.}
\label{fig:0cntConFR}
\end{figure*}
\begin{figure*}
\includegraphics[width=7.2cm]{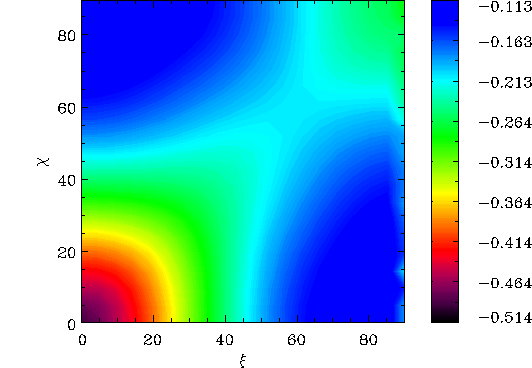}
\includegraphics[width=7.2cm]{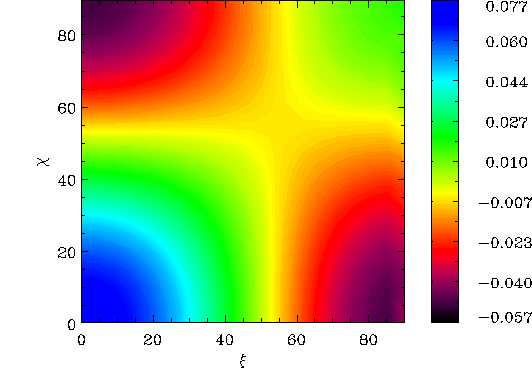}
\caption{Same as in Figure \ref{fig:0cntConFR} but for the fixed-ion crustal
emission model.}
\label{fig:0cntConFX}
\end{figure*}

Figure \ref{fig:0cntAt} shows the contour plots
of the phase-averaged polarization fraction 
 in the ($\chi$, $\xi$) plane, for the X--ray 
($0.12-0.39$ keV) and  optical (B-filter) bands. 
In both cases the phase-averaged polarization 
degree is significantly high. Like the results already obtained for the
viewing geometry used in Figure \ref{fig:0lc},  
the phase-averaged polarization degree  in the X--ray band is $\sim
99\%$, and
that in the optical band is only slightly lower, $\sim 87\% $.

It is important to stress that these plots (and the analogous ones in
\S\ref{rescon}) show the polarization
degree as computed by using the definition given in equation~(\ref{polint}),
i.e. considering the difference in the radiative flux carried by the
two modes when radiation reaches infinity, and repeating
the calculation at different spin phases. Although we take
into account for relativistic ray bending (i.e. for the fact that the
emitting area which is into view is larger than a hemisphere),
a proper calculation of the polarization observables is based on the Stokes
parameters and must account for both QED effects in the magnetized vacuum
through which photons propagate and the rotation of the plane normal to the
photon wavevector in a varying magnetic field (see \S\ref{vac}),
effects that are not accounted for in the plots of Figure \ref{fig:0cntAt}.
We therefore refer to this quantity as the ``intrinsic'' degree of linear
polarization, to distinguish it from the observed one, which is discussed later
on (see section 4). We remark that both the ``intrinsic'' and the observed
degree of polarization are evaluated at infinity, and a comparison of the two
quantities may be of help in understanding how QED and geometrical effects
influence the polarization observables.

\subsubsection{Condensed surface}
\label{rescon}


The same approach described in the previous subsection was used to compute the
(phase-dependent) spectrum and the intrinsic polarization fraction
for a bare INSs with a condensed surface. In this case the calculation was
repeated twice, by assuming either ``free'' or 
``fixed'' ions. Since we adopt the approximated analytical expressions by
\cite{potekhin12} for the emissivities, no interpolations were required,
contrary to the case of the atmosphere.
{The fitting formulae, however, cover the range $0\leq\theta_\mathrm B\leq
\pi/2$. In order to take into account the emission from the southern
magnetic hemisphere of the star, where $\mathbf B$ ``enters'' into the surface
and $\pi/2< \theta_\mathrm B\leq \pi$, the emissivities are calculated by
replacing $\cos\theta_\mathrm B$ with $-\cos\theta_\mathrm B$ and
$\cos\phi_\mathrm k$ with $-\cos\phi_\mathrm k$ (A. Potekhin, private
communication).} Results are reported in Figure \ref{fig:0lcCon}, where  
the X--ray lightcurve and the phase dependent polarization degree are shown for 
$\chi=90^\circ$ and $\xi\sim 18^\circ$, which is again compatible with a
pulsed
fraction of $\sim 
1\%$.\footnote{We find that, when using 
a crustal emission model, the domain of viewing angles for  
which the X--ray pulsed fraction is $\sim 1\%$ are not too different from 
those obtained using an atmospheric model.}
The corresponding contour plots for the
X--ray and optical phase-averaged polarization fraction are shown in
Figure \ref{fig:0cntConFR}
(free ions) and Figure \ref{fig:0cntConFX} (fixed ions), respectively.
{In the case of free ions, the phase-averaged polarization degree in the
X--ray band is always less  than $\sim 6\%$ either when the emission is
dominated by  ordinary (negative values) or extraordinary photons (positive
values, see Figure \ref{fig:0cntConFR}, right
panel), while in the  optical band (Figure \ref{fig:0cntConFR}, left panel)
ordinary photons are predominant, with a maximum polarization degree $\sim 30\%$
for particular viewing geometries. In the case of fixed ions expectations are
similar in the X--ray band, where the phase-averaged polarization degree is
always less than $\sim 8\%$. However, in the optical band,  the polarization
degree can be slightly larger, up to $\sim 50\%$ (Figure
\ref{fig:0cntConFX}, left panel)}, for the most favorable viewing geometries.

\subsection{Stokes parameters and vacuum polarization} 
\label{vac}

\begin{figure*}
\includegraphics[width=5.9cm]{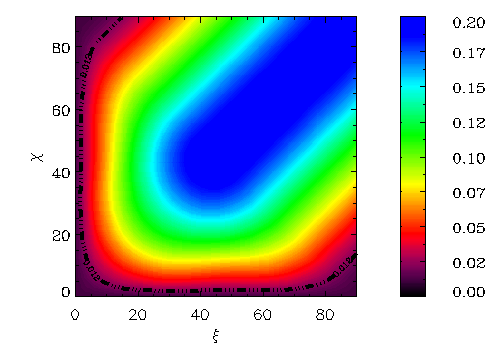}
\includegraphics[width=5.8cm]{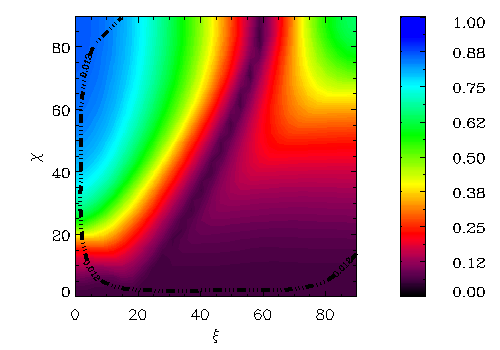}
\includegraphics[width=5.8cm]{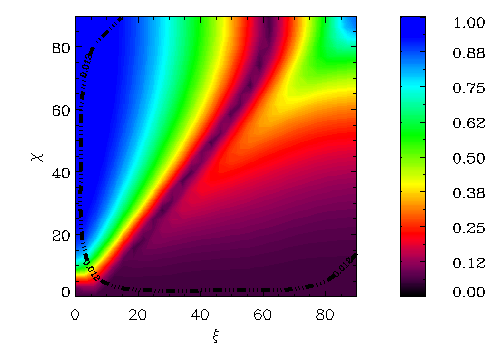}
\caption{
Contour plots for the X--ray pulsed fraction, phase-averaged polarization
fraction in the optical (B filter) 
band and phase-averaged polarization fraction in the X--ray band 
 in the ($\chi$,$\xi$) planes (panels from left to right respectively) for the
gaseous atmosphere model. 
 All polarization observables are computed by the expressions of the Stokes
parameters (see \S~\ref{vac}). 
 The black curve in each panel corresponds to the observed X--ray pulsed
fraction of RX J1856, $\sim 1\%$.}
\label{fig:cntAtm}
\end{figure*}
A strong magnetic field can modify the properties of the vacuum
outside the NS. In particular, due to QED effects, photons can
temporarily convert into electron-positron pairs, and those
virtual pairs modify the dielectric and the magnetic permeability
tensors of the vacuum. This affects the direction of the photon
electric field and, in turn, influences the polarization fraction
as observed at infinity \cite[][see also \citealt{hardlai06}]{heyl00,heyl02}. As
a linearly polarized
electromagnetic wave propagates in the magnetized vacuum close to the star, the
direction of the
electric field changes on a spatial scale much shorter than that over which
$\mathbf B$ varies.
This implies that up to the adiabatic (or polarization-limiting)
radius\footnote{Equation (\ref{adiarad}) holds for a dipole field.}
\begin{equation}
\label{adiarad}
r_\mathrm a \simeq 4.8\left(\frac{B_\mathrm p}{10^{11}\ \mathrm G}\right)^{2/5}
\left(\frac{E}{1\ \mathrm{keV}}\right)^{1/5} R_\mathrm{NS} \, ,
\end{equation}
a photon will keep the polarization mode (either X or O) in which it was
emitted.
Around $r_\mathrm a$, the coupling between $\mathbf B$ and the wave electric
field weakens, until for $r\gg r_\mathrm a$, the direction of 
the electric field freezes \cite[][paper I]{heyl03,ferndav11}.

The evolution of the wave electric field can be followed by solving the (linearized)
wave equation in the magnetized vacuum around the star along each 
photon trajectory. However, as shown in paper I, the main effects of
vacuum polarization can be cought using a simpler approach in which adiabatic
propagation (i.e. mode locking) is assumed up to $r_\mathrm a$, while the
electric field direction is constant (and modes change) for  $r > r_\mathrm a$.
In this approach
the polarization properties are determined by the direction of the magnetic
field at $r_\mathrm a$, in addition to the intrinsic polarization degree at the
surface.

\begin{figure*}
\includegraphics[width=5.8cm]{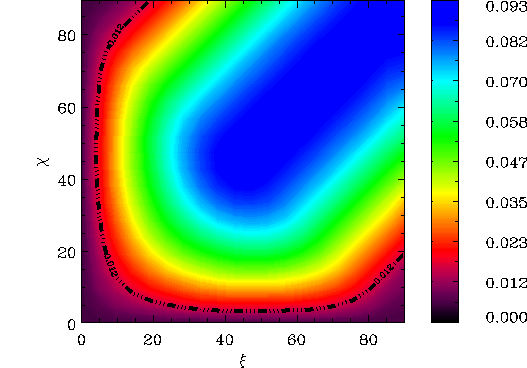}
\includegraphics[width=5.8cm]{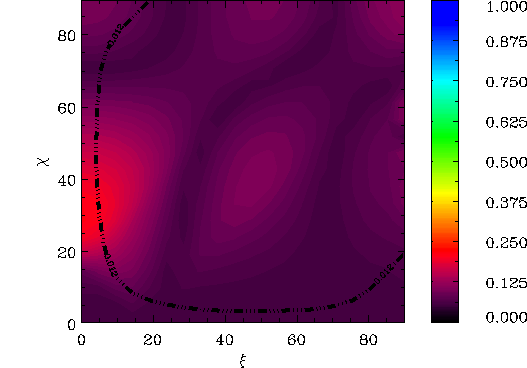}
\includegraphics[width=5.8cm]{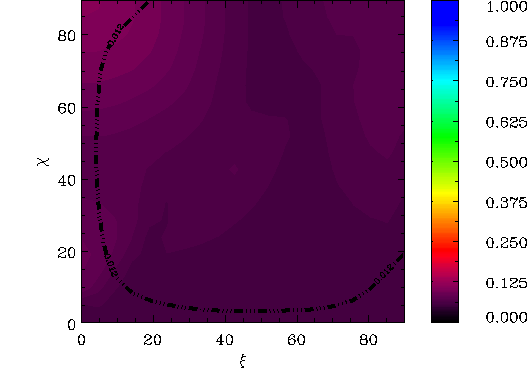}
\caption{Same as Figure \ref{fig:cntAtm} but for the free-ion crustal emission
model.} 
\label{fig:cntCfree}
\end{figure*}

\begin{figure*}
\includegraphics[width=5.8cm]{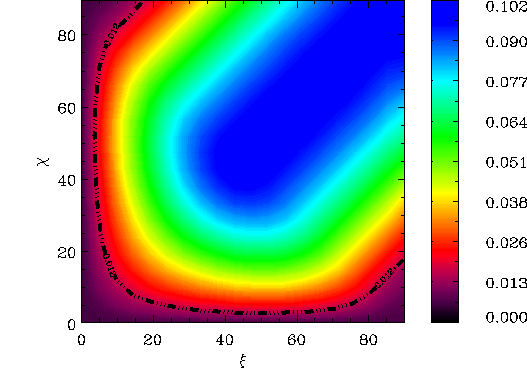}
\includegraphics[width=5.8cm]{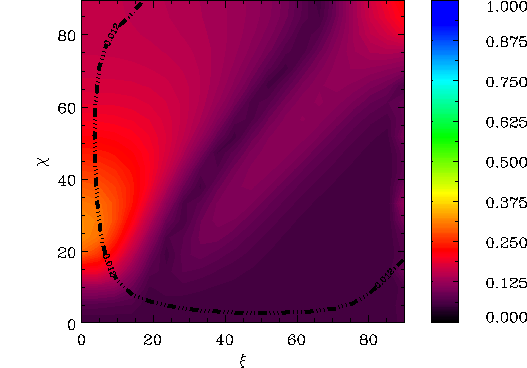}
\includegraphics[width=5.8cm]{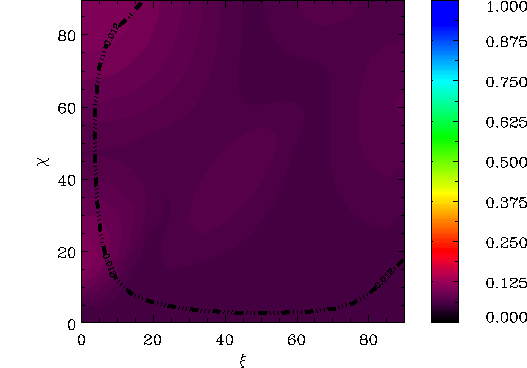}
\caption{
Same as Figure \ref{fig:cntAtm} but for the fixed-ion crustal emission model.} 
\label{fig:cntCfix}
\end{figure*}

Since the X and O modes are defined according to the direction of the
wave electric field with respect to the plane spanned by the magnetic field
$\mathbf B$ and the wavevector  $\mathbf k$, the Stokes parameters of photons
crossing $r_\mathrm a$ at different positions are, in general, referred to
different coordinate systems. While the $z_i'$ axes coincide with the LOS (i.e.
with $\boldsymbol\ell$), the two axes, $x_i',\, y_i'$, in the plane orthogonal
to $\boldsymbol\ell$ will  be different for each photon trajectory, because
$\mathbf B$ changes its direction over the sphere of radius $r_\mathrm a$. In
order to derive the polarization observables, as detected by a distant
instrument, the Stokes parameters must be referred to the same fixed direction
in the  plane perpendicular to the LOS, $\mathbf u$. This is done by rotating
the Stokes parameters by an angle $\alpha_i=\arccos\mathbf u\cdot {\mathbf x}_i$
(for the choice of the sign of $\alpha_i$ see, Paper I)
\begin{eqnarray}
I_i&=&\bar I_i \nonumber\\
Q_i&=&\bar Q_i \cos(2\alpha_i) + \bar U_i \sin(2\alpha_i)\\
U_i&=&\bar U_i \cos(2\alpha_i) - \bar Q_i \sin(2\alpha_i)\, .\nonumber
\end{eqnarray}

In a strong magnetic field, each photon is 100$\%$ polarized
either in the X-mode or O-mode.
This is conveniently expressed in terms of the (normalized)
Stokes parameters of each
photon as  $\bar I_i = 1$, $\bar Q_i = \pm 1$ (for X-mode and O-mode photons) and $\bar U_i=
0$ (see paper I).
The collective Stokes parameters, i.e. those for the entire radiation field, are
simply the sum of the individual parameters.
This is immediately generalized to a continuum photon
distribution following the same approach as in equation (\ref{flux}) 
\begin{eqnarray}
\label{stokesflux}
I&=&\int_0^{2\pi} d\Phi_\mathrm s\int_0^1 du^2(n_\mathrm X+n_\mathrm
O)\nonumber\\
Q&=&\int_0^{2\pi} d\Phi_\mathrm s\int_0^1 du^2(n_\mathrm X-n_\mathrm
O)\cos(2\alpha)\\
U&=&\int_0^{2\pi} d\Phi_\mathrm s \int_0^1 du^2(n_\mathrm O-n_\mathrm
X)\sin(2\alpha)\nonumber\,,
\label{eq:intOX}
\end{eqnarray}
where $n_\mathrm{O,X}=I_\mathrm{O,X}/E$, and $Q,\, U$ are the fluxes of the
Stokes parameters; here $I$ is proportional to the total number flux and in 
(\ref{stokesflux}) the constant
factor appearing in front of the integral (see equation [\ref{flux}]) has been
dropped. The explicit expression for $\alpha$ as a function of  $\Theta_\mathrm
s$, 
$\Phi_\mathrm s$, $\xi$, $\chi$, the phase $\gamma$ and the photon energy, $E$
was derived in paper I.
Finally, the observed polarization fraction and polarization angle are given by
the usual expressions
\begin{eqnarray}
\label{polobs}
\Pi_\mathrm L  &=& \frac{\sqrt{Q^2+U^2}}{I}\\
\label{polobs1}
\chi_\mathrm P &=& \frac{1}{2}\arctan \left( \frac{U}{Q} \right) \, .
\end{eqnarray}

\section{The observed polarization signal}
\label{obspol}

With the method described in \S~\ref{vac}, we can determine the polarization of the radiation as 
measured by a distant observer for any given viewing configuration. In particular, we 
compute and compare the  X--ray pulsed fraction, the phase-averaged degree of
polarization and polarization angle in 
the X--ray and in the optical bands as 
functions of the two geometrical angles $\chi$  and $\xi$. 
 Here, all the calculations are performed by
assuming that the polarimeter reference frame is aligned with the ``fixed''
one on the NS.  We should notice that the choice of the direction of the polarimeter  reference
frame  with respect to the ``fixed'' reference frame of the
neutron star has no effect on  the polarization fraction, but it affects
the polarization angle (see \S~\ref{vac} and paper I for details).

The computed X--ray pulsed fractions are  quite similar for both the atmosphere
(Figure \ref{fig:cntAtm}, 
left panel) and condensed surface (Figure \ref{fig:cntCfree} and
\ref{fig:cntCfix}, left panels).
Particularly, the $1\%$ X--ray pulsed fraction observed in RX J1856 does not
impose a strong constraint on the viewing
geometry\footnote{ Using a magnetized model atmosphere,  \citet{ho07a}
constrained the viewing geometry of RX J1856 to  $<6^\circ$  for one angle and
$\approx 20^\circ-45^\circ$ for the other. Our ranges for the viewing angles are
compatible but less restrictive. The discrepancy may be
due to the different choice of mass, radius and temperature
which are $M=1.4 \rm{M}_\odot$,
$R=14$~km (note that the value of the radius was based on a different distance
estimate) and $T_p = 7\times 10^5$~K (at the magnetic pole) and $T_{eq}=
4\times10^5$~K (at the magnetic equator)  in \citet{ho07a}.} of
the NS ($\chi$ and
$\xi$ angles). However, it imposes (to a
minor extent) a constraint to the polarization observables. So, for comparison
and completeness we also keep this information in the contour plots of
phase-averaged polarization fraction and polarization angle.

Figure \ref{fig:cntAtm} shows our results for the case of  magnetized
atmospheric model. First, we note that the range of viewing
angles in which the polarization fraction is substantial (and potentially
detectable) is approximately given by  $\chi>\xi$. Viewing geometries near $\chi=90^\circ$, $\xi=0^\circ$ 
 or $\chi=90^\circ$, $\xi=90^\circ$, 
 which correspond to aligned and orthogonal rotators respectively, both seen perpendicularly to the spin axis, are
particularly
favorable for observing a high phase-averaged  polarization fraction. In
particular,  for $\chi=90^\circ$ and $\xi=0^\circ$   the phase-averaged
polarization fraction
can reach $\sim 80\%$ in the optical band, and be even larger, up to $\sim
90\%$, in the X--ray band.

Figure  \ref{fig:cntCfree} and  \ref{fig:cntCfix} show the case of
a  condensed surface in the two limits, free and
fixed ions, respectively. The results are noticeably different with respect to the atmospheric model.
{In fact, for free ions, if we consider for example viewing
geometries close to $\chi=40^\circ$ and $\xi=0^\circ$   the phase-averaged
polarization
fraction can still be as large as $\sim 20\%$ in the optical band but it is
substantially 
reduced in the X--ray band. In the case of fixed ions, for similar viewing
geometries,  we expect a phase-averaged polarization fraction of  $\sim 35\%$ in the
optical band while, in the X--ray band, the polarization is
vanishingly small for all viewing angles.}

As noticed in  \citet{ferndav11} and in paper I, the
phase-averaged polarization fraction is expected to be small for
$\chi<\xi$, due to a combination of both QED effects and the frame
rotation of the Stokes parameters which is needed in presence of a varying
magnetic  field over the emitting region.
In paper I, the emission from a NS is computed for a dipolar
magnetic field distribution and $100\%$ polarized blackbody emission, and
it is found that in almost the entire region $\chi<\xi$ the phase-averaged
polarization fraction is roughly zero, consistently with the present results.
The effects of the choice of the emission model become important for viewing angles
$\chi>\xi$.
In particular, for a magnetized atmosphere, the
highest phase-averaged polarization fraction is attained in the
region near $\chi=90^\circ$ and $\xi=0^\circ$. This is because: i) under this
hypothesis
an  ``intrinsic'' polarization fraction (see \S~\ref{resatmo})
as high as $\sim 99\%$ is expected, and ii) in the
case of an aligned rotator there is virtually no differential effect due
to the rotation of the Stokes parameters at the adiabatic radius (that
tends to reduce the observed polarization degree).

{The situation is different for the condensed surface emission.
In the optical band, in fact, we expect a maximum of the ``intrinsic'' 
polarization fraction as high as $\sim 30\%$ ($\sim 50\%$) for the case 
of free (fixed) ions at viewing angles $\chi \sim \xi \sim 0^\circ$ (see 
\S~\ref{rescon}). However, for this viewing geometry, the depolarizing effects
of the Stokes parameter rotation are stronger because the $\alpha$ angle
distribution assumes symmetrically all the values in the range $[0,2\pi]$,
cancelling the original ``intrinsic'' polarization at infinity. As shown in
Figures \ref{fig:cntCfree} and \ref{fig:cntCfix}, central panels, a significant
polarization degree, as high as $\sim 20\%$ ($\sim 35\%$) for the case of free
(fixed) ions, is present only for sligthly greater values of $\xi$, before
decreasing again according to the behaviors shown in the plots of Figures
\ref{fig:0cntConFR} and \ref{fig:0cntConFX} (left panels). On the other
hand, in the X--ray band,} ordinary and extraordinary waves are expected to have
similar reflected amplitudes: the ``intrinsic'' polarization fraction is
therefore substantially reduced and as well the observed ones (at all viewing
angles).

\begin{figure*}
\includegraphics[width=7.cm]{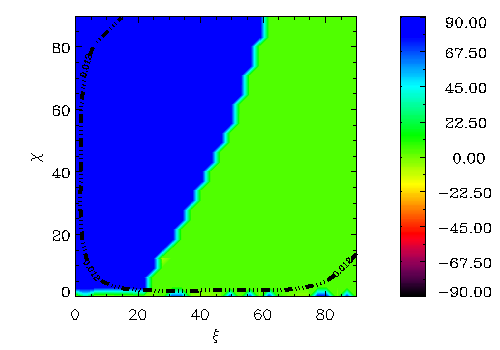}
\includegraphics[width=7.cm]{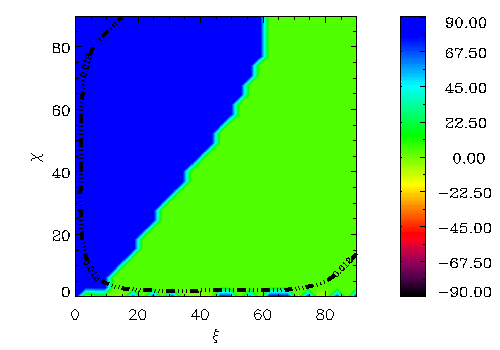}
\caption{
Contour plots of the phase-averaged polarization angles in the ($\chi$, $\xi$)
plane 
  for the gaseous atmosphere emission model, with left and right panels 
  corresponding to the optical (B filter)  
  and the X--ray bands respectively.}
\label{fig:AtmAng}
\end{figure*}

\begin{figure*}
\includegraphics[width=7.cm]{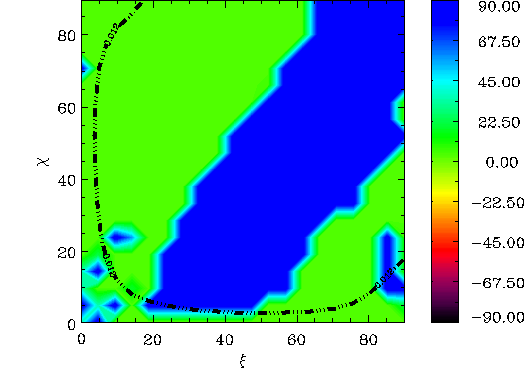}
\includegraphics[width=7.cm]{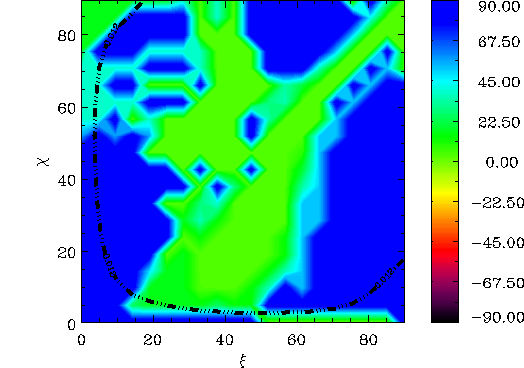}
\caption{Same as Figure \ref{fig:AtmAng} but for the fixed-ion crustal emission
model.}
\label{fig:CfixAng}
\end{figure*}

Figure \ref{fig:AtmAng} shows the phase-averaged polarization angle for
the atmospheric emission in  the optical and X--ray bands.  The computed
quantity is nearly constant in two regions of viewing angles, for which it is $\sim
90^\circ$ and $\sim 0^\circ$.
This occurs also for the condensed surface models
({see Figure \ref{fig:CfixAng} in the case of fixed ions});
however, the main difference is that in these cases the expectation are
{somehow}
reversed in the two bands.  In fact, {by considering the region of 
viewing angles in which the phase-averaged polarization fraction is
detectable, 
$\chi\sim30^\circ$ and $\xi\sim 0^\circ$,} we can see that in the case of fixed
ions the expected phase-averaged
polarization angle in the optical is $\chi_\mathrm P \sim 0^\circ$, while the X--ray band 
this is  $\chi_\mathrm P\sim 90^\circ$.

 Again, this behavior can be understood in terms of QED effects. 
The polarization angle reflects  the global direction of the electric field
distribution of the radiation, which  in turn depends on the direction of the 
magnetic field at the adiabatic radius.
Then,  the observed phase-averaged
polarization angle should reflect the ``phase-averaged'' direction of the
magnetic field at the adiabatic radius, which for viewing angles $\chi>\xi$
is approximately parallel and for $\chi<\xi$ is approximately
perpendicular to the spin axis. As a consequence, if the observed
radiation is dominated by extraordinary photons, then for $\chi>\xi$ the
``average'' observed direction of the photon electric field is
perpendicular to the spin axis, and the phase-averaged polarization angle is
$\sim 90^\circ$, in agreement with our expectations for the case of the
atmosphere model in both, the optical and the X--ray band (Figure
\ref{fig:AtmAng}, left and right panel, respectively). Here, we should
notice that the association  between the normal modes and the phase-averaged
polarization fraction is possible because we already set the coordinate system
of the polarimeter aligned with the ``fixed'' coordinate system of the NS. 
However, in general  the reference frames of the polarimeter 
and the ``fixed'' one of the NS are expected to be disaligned.

For condensed surface emission (in both approximations, fixed and
free ions), whereas in the optical band the emitted radiation is dominated
by ordinary photons { (see Figure \ref{fig:0cntConFX}, left panel), in the
X--ray band the two modes have similar
intensities, with the emitted radiation being slightly dominated by
extraordinary photons for fixed ions (Figure \ref{fig:0cntConFX}, right
panel)}. 
As a consequence, in the optical band and
for viewing angles $\chi>\xi$ we expect that the phase-averaged
direction of the photon electric field is parallel to the spin axis, and
thus the phase-averaged polarization angle is $\sim 0^\circ$. On the
contrary, in the X--ray band and again for viewing angles
$\chi>\xi$ the phase-averaged direction of the photon electric field is
perpendicular to the spin axis, and therefore the expected phase-averaged
polarization angle is $\sim 90^\circ$. {However, the behavior of
the polarization angle in the X--rays is quite irregular, due to the fact that
the emissivities of the two modes in this band are similar to each other. So,
the polarization angle present jumps by 90$^\circ$, which 
arise because of an even slight predominance of O over  X photons or
conversely. For the same reason, we do not show the contour plots in the case of
free ions, since 
the corresponding polarization fraction in the two energy bands is even lower 
than that of the fixed-ions case; hence, the phase-averaged polarization angle
behavior for free ions present even more noisy patterns.} 

The main conclusion is that, by measuring the phase-averaged polarization
observables in optical {\it and} X--ray
bands it is potentially possible to discriminate between atmospheric and
crustal emission. The most favorable geometries are those with viewing
angles $\chi>\xi$, for which the expected phase-averaged polarization
fraction is substantial. If emission is atmospheric, we expect a high
phase-averaged polarization fraction in both, optical and X--ray band
(although slightly lower in the optical). Whereas, if emission originates from a
condensed surface, the phase-averaged polarization fraction should be
{modest in the optical, with an almost unpolarized signal in
the X--ray band}.

At the same time,  the phase-averaged polarization angle for atmospheric emission
is expected to be the same in the optical and X--ray band. On the
contrary, for crustal emission the angle measured in the two bands is
expected to be different by $\sim 90^\circ$ (although this latter
consideration is just a theoretical expectation, since realistically the measure
of the angle in the X--ray band can not be performed due to low degree of 
polarization in the signal).

\section{ Discussion and Conclusions}
\label{disc}

We studied the polarization properties of the thermal emission from RX
J1856 considering  different emission models: a NS with a magnetized
atmosphere or a condensed surface. The effects of vacuum
polarization in the propagation of the radiation in the NS magnetosphere and
the rotation of the Stokes parameters were accounted for.
Using a ray tracing method and assuming a dipole magnetic field,  we derived
the polarization observables  for different viewing angles $\chi$ and
$\xi$.
We found that the phase-averaged polarization fraction can be substantially large
for viewing angles $\chi > \xi$, which is consistent with previous works (see
paper I and references therein). For these viewing angles, in the case of an
atmosphere, we found  that a) the phase-averaged polarization fraction in the 
optical band is expected to be lower than in the X--ray band and b) the
phase-averaged polarization angle in the optical is the same that in the X--ray
band. In the case of condensed surface, we found that a) the
phase-averaged polarization fraction in the optical band is substantially higher
than in  the X--ray band and b) the phase-averaged polarization angle in the
optical
band {is generally shifted by 90$^\circ$} with respect to that  in the
X--ray band. 
Therefore, by combining optical and X--ray  observations of the polarized
emission from  RX J1856, it is possible to determine if this XDINS has an
atmosphere or a condensed surface.

Our treatment of the surface emission from RX J1856 relied
on a number of assumptions/simplifications which are discussed in
more detail below. In this respect we stress that our main goal
has been to assess to which extent polarization measurements at
optical and X--ray energies are effective in discriminating among
different emission models, rather than to derive theoretical
predictions to be matched with current observations (e.g. through
spectral fitting). It is in this spirit that we deliberately chose
to restrict to a simple treatment of the emission models we
employed, in particular for the atmospheric model\footnote{Our
treatment of the condensed surface emission is state of the art,
although the presence of a thin, H layer on top of the star was
not included.}.

A major simplification we introduced is that of pure H composition
and complete ionization.
 For the low surface temperature ($\sim
60$ eV) and strong magnetic field ($\sim 10^{13}$ G) of RX J1856, the neutral
fraction of H atoms is expected to be $\approx 0.01$--$0.1$ for typical
atmospheric densities \cite[][]{potcha04}, so that 
opacities are affected. H atmospheres with partial ionization have been
presented e.g. by \cite{pot04} and \citet[see also
\citealt{potekhin14}]{suleim09}. The major difference with respect to fully
ionized models is the appearance of spectral features related to atomic
transitions. These features, however, are mainly confined to 
far UV--soft X--ray range ($\la 0.2$ keV), and fully ionized models give a
reasonable description of the spectra at X--ray/optical energies. 
Moreover, the features are strongly smeared out when the contributions from
different surface patches (each with a different $T$ and $\mathbf B$) are
summed together to obtain the spectrum at infinity \cite[][see again also
\citealt{potekhin14}]{ho08}, similarly to what occurs to the proton cyclotron
line \cite[][]{zane01}. 

As noticed by \citet[see also \citealt{vanadel06}]{holai03}, in
the atmosphere around a strongly magnetized neutron star vacuum
polarization can induce a Mikheyev-Smirnov-Wolfenstein like
resonance across which a photon may convert from one mode into the
other, with significant changes in the opacities and polarization.
While for $B\la 10^{13}$~G this is not going to change the
emission spectrum, it still can significantly affect the
polarization pattern at least at certain energies. For a photon of
energy $E$, the vacuum resonance occurs when the vacuum and plasma
contributions to the dielectric tensor become comparable, i.e.
where $\rho = \rho_V \approx 0.96 \times 10^{5} Y_e^{-1} (E/ 1\
\mathrm{keV})^2 (B/10^{14}\mathrm{G})^2 f^{-2}$ g cm$^{-3}$, where
$Y_\mathrm e = Z/A$ ($Z$ and $A$ are the atomic and mass number of
the ions) and $f\sim 1$ is a slowly varying function of $B$. Near
the vacuum resonance, the probability of mode conversion is given
by $1- P_\mathrm{jump} = 1- \exp \left [ - \pi\left
(E/E_\mathrm{ad} \right )^3/2\right ]$, where $E_\mathrm{ad}$
depends on the photon energy, on $B$ and on the angle between the
photon direction and $\mathbf B$ \cite[$\theta_\mathrm{Bk}$,][see
in particular their eq. 3]{vanadel06}. For $B \la 10^{13}$~G (as
in the case discussed in this work), it is $\rho_\mathrm V <
10^{-3}$~g/cm$^{-3}$, i.e. the vacuum resonance is well outside
the photospheres of both the ordinary and the extraordinary mode.
Moreover, the inequality $E<E_\mathrm{ad}$ is satisfied for all
photon energies $\la 1$ keV, unless radiation is propagating
nearly along the magnetic field direction ($\tan
\theta_\mathrm{Bk}\la 0.1$). For this reason our assumption of
``no mode conversion'' at the vacuum resonance, which is
equivalent to assume $E\ll E_\mathrm{ad}$ (or $P_\mathrm{jump}=1$)
for all photons, appears reasonable. Further narrow features due
to mode collapse and spin-flip transitions are expected very near
the broad proton cyclotron resonance \cite[][see  also
\citealt{mel81} for the case of electrons]{holai03, zane00}. In
the absence to a complete description of the dielectric tensor in a
electrons+ions+vacuum plasma we assumed as a working hypothesis no
mode conversion at this frequency.

The present analysis can be extended to other XDINSs as well. 
The narrow range of surface temperatures  inferred from the spectra of XDINSs,
$T\sim 50 -100$~eV may be important to determine the state of the
surface, but it should not have an important effect on the properties of the
observable polarization.
A significant difference on the polarization properties of XDINSs  may be
introduced if we consider different magnetic field configurations (see
paper I for the example of a twisted magnetic field).  However, in general,
XDINSs share similar magnetic field
configuration, i.e., external dipole magnetic field, and there is no
observational evidence for  multipolar components or
twisted
magnetic fields (such as those that may be present in magnetars, see
\citealt{turolla15}). This is supported by
the good agreement between the magnetic fields derived from timing
properties and those inferred from the absorption lines
(assuming that they are caused by proton cyclotron resonance, see
\citealt{turolla09}), and the absence of non-thermal emission  that
may be linked to the presence of a  twist in the external magnetic field. 
However, in this respect RX J0720.4-3125 may be an exception. For this XDINS, an
absorption feature that is energy-dependent and
phase-dependent has been  recently reported \citep{borghese15}. If
this feature is caused by proton cyclotron resonance, then  it would be
compatible
with the presence of a  multipolar component confined  very close to the NS
surface and consistent with a magnetic field $B=10^{14}$~G.  
The effect of this component on the polarization properties of the radiation
has not been assessed, but certainly it can be studied using the method
developed in paper I and the emission models here discussed.

\section*{Acknowledgments}
We would like to thank the referee, dr. A. Y. Potekhin, for his constructive
criticism and helpful comments which greatly improved a previous version of the
paper. 
D. Gonz\'alez Caniulef aknowledges financial support from the RAS and the 
University of Padova for funding his visit at the
Department of Physics and Astronomy, University of Padova, 
where part of this investigation was carried out.
He also acknowledges the financial support by CONICYT through a 
``Becas Chile'' fellowship No. 72150555.  The work of R. Turolla 
is partially supported by an INAF PRIN grant.

\end{document}